\documentclass[reviewcopy]{elsart}

\usepackage[english]{babel}
\usepackage[latin1]{inputenc}
\usepackage[T1]{fontenc}

\usepackage{color,graphicx,graphics,lineno}
\usepackage{float}
\usepackage{amsfonts}
\usepackage{amssymb}
\usepackage{amsmath}
\usepackage{bm}
\usepackage{bbm}
\usepackage{natbib}

 {\everymath{\displaystyle\everymath{}}\array}%
 {\endarray}
\pagewiselinenumbers%
\begin{document}
\begin{frontmatter}
\vspace{-1.5cm}
\title{Spatio-temporal Functional Regression on Paleo-ecological Data}
\author{Liliane Bel\corauthref{cor}},
\corauth[cor]{Corresponding author.} \ead{Liliane.Bel@agroparistech.fr}
\address{UMR 518 AgroParisTech/INRA,16, rue Claude
Bernard - 75231 Paris Cedex 05 }
\author{Avner Bar-Hen},
\address{Universit\'{e} Ren\'{e} Descartes,  MAP5, 45 rue des Saints P\`{e}res, 75270 Paris cedex 06}
\author{Rachid Cheddadi},
\address{ISEM, case postale
61, CNRS UMR 5554, 34095 Montpellier, France}
\author{R\'{e}my Petit},
\address{UMR 1202 INRA , 69 route d'Arcachon
33612 Cestas Cedex, France}
\begin{abstract}
The influence of climate on biodiversity is an important ecological
question. Various theories try to link climate change to allelic
richness and therefore to predict the impact of global warming on
genetic diversity. We  model the relationship between genetic
diversity in the European beech forests and  curves of temperature
and precipitation reconstructed from pollen databases. Our model
links the genetic measure to the climate curves through a linear
functional regression. The interaction in climate variables is
assumed to be bilinear. Since the data are georeferenced, our
methodology accounts for the spatial dependence among the
observations. The practical issues of these extensions are
discussed.
\end{abstract}
\end{frontmatter}

\section{Introduction}
Climate records show that the earth has recorded a succession of
periods of major warming and cooling at different time windows and
scales \cite{Dahl-Jensen, Seierstad}.  During the last post-glacial
period (18000 years before the present), Europe recorded a 15°C to
20°C warming depending on the area. At the same period there was an
expansion of all forest biomes and an upward movement of the
tree-lines that reached an altitude 300 m higher than today.
Although there is a wealth of paleodata  and detailed climate
reconstruction for the Holocene period, we still lack some knowledge
as to how the warming was recorded and what the vegetation feedbacks
were that  affected local or regional past climates. Various
theories try to link climate change to allelic richness and
therefore to predict the impact of global warming on genetic
diversity.

In the recent literature there have been a lot of  theoretical
results for regression models with functional data. Based on this
framework, we used a linear functional model to
 model the relationship between genetic diversity in European beech forests (represented by a
positive number) and  curves of temperature and precipitation
reconstructed from the past. The classical functional regression
model has been extended in two ways to account for  our specific
problem. First, as the effects of temperature and precipitation are
far from independent we included an interaction term in our model.
This interaction term appears as a  bilinear function of the two
 predictors. Second, since we have spatial data
there is dependence among the observations. To take into account
with dependence the covariance matrix of the residuals is estimated
in a spatial framework and plugged into
 generalized least-squares to estimate the parameters
of the model. The practical difficulties of these extensions will be
discussed.

In Section 2, we present the genetic and climate data. The
functional regression model is studied in Section 3. Results are
presented and discussed in Section~4 and concluding remarks are
given in Section~5.

\section{Data}

Pollen records are important proxies for the reconstruction of
climate parameters since variations in the pollen assemblages mainly
respond to climate changes. Based on the fossil and surface pollen
data from pollen databases, we used modern analogue technique (MAT)
to reconstruct climate variables. Climate reconstruction is
accomplished by matching fossil biological assemblages to recently
deposited (modern) pollen assemblages for which climate properties
are known. The relatedness of fossil and modern assemblages is
usually measured using a distance metric that rescales
multidimensional species assemblages into a single measure of
dissimilarity. The distance-metric method is widely used among
paleoecologists and paleoceanographers \cite{Guiot}. Temperature and
precipitation were reconstructed at 216 locations from the present
back to a variable date depending on  available data. The pollen
dataset was used to reconstruct climate variables, throughout Europe
for the last 15 000 years of the Quaternary. Due to the methodology,
each climate curve is sampled at irregular times for each location.

Genetic diversities were measured from variation at 12 polymorphic
isozyme loci in European beech ({\em Fagus sylvatica} L.) forests
based on an extensive sample of 389 populations distributed
throughout the species range. Based on these data, various indices
of diversity can be computed. They mainly characterize within or
between population diversity. In this article, we focus on the
$H$~index, the probability that two alleles sampled at random are
different. This parameter is a good indication of gene diversity
\cite{comps}.

The two datasets were collected independently and their locations do
not coincide.

\section{Functional Regression}
The functional linear regression model with functional or real
response has been the focus of various investigations
\cite{Cardot,Fan, Faraway,Ramsay}. We want to estimate the link
between the real random response $y_i=d(s_i)$, the diversity at site
$s_i$ and $(\theta_i(t),\pi_i(t) )_{t>0}$ the temperature and
precipitation functions at site $s_i$. There are two points to
consider for the modeling: (i) functional linear models need to be
extended to incorporate interaction between climate functions; (ii)
since we have spatial data, observations cannot be considered as
independent and we also need to extend functional modeling to
account for spatial correlation.

We assume that the temperature and precipitation functions are
square integrable random functions defined on some real compact set
$[0,T]$. The very general model can be written as:
\[
  Y = f((\theta(t),\pi(t))_{T>t>0}) + \varepsilon
\]

$f$ is an unknown functional from $L^2([0,T]) \times L^2([0,T])$ to
$\mathbb{R}$ and $\varepsilon$ is a spatial stationary random field
with correlation function $\rho(.)$.

We assume here that the functional $f$ may be written as the sum of
linear terms in $\theta(t)$ and $\pi(t)$ and a bilinear term
modeling the interaction

\begin{eqnarray*}
  f(\theta,\pi)& =&\mu+\int_{[0,T]}A(t)\theta(t)\mbox{d}t+\int_{[0,T]}B(t)\pi(t)\mbox{d}t+
  \int \! \!  \int_{[0,T]^2} C(t,u)\theta(t)\pi(u)dudt\\
 & =& \mu+\langle A ; \theta\rangle+\langle B ; \pi\rangle+\langle C\theta ; \pi\rangle
\end{eqnarray*}

 by the Riesz representation of linear and bilinear forms.

$A$ and $B$ are in $L^2([0,T])$ and $C$ is a kernel of $L^2([0,T])$.

Let $(e_k)_{k>0}$ be an orthonormal basis of $L^2([0,T])$. Expanding
all functions on this basis we get
\[
\theta_i(t)= \sum_{k=1}^{+\infty} \alpha_k^ie_k(t)\quad \pi_i(t)=
\sum_{k=1}^{+\infty} \beta_k^ie_k(t)\]
\[ \quad A(t) = \sum_{k=1}^{+\infty} a_ke_k(t) \quad B(t) = \sum_{k=1}^{+\infty} b_k e_k(t)
\quad C(t,u) = \sum_{k,\ell=1}^{+\infty} c_{k \ell}e_k(t)e_\ell(u)
\]
and
\[
  y_i = \mu + \sum_{k=1}^{+\infty}a_k\alpha_k^i + \sum_{k=1}^{+\infty} b_k \beta_k^i  +
  \sum_{k,\ell=1}^{+\infty} c_{k \ell}\alpha_k^i\beta^i_\ell +
  \varepsilon_i
\]

If the sums are truncated at $k=\ell=K$ the problem results in a
linear regression $Y = \mu + X\phi + \varepsilon$ with spatially
correlated residuals with

\[
 X = \begin{pmatrix}
 \alpha_1^1 & \ldots & \alpha_K^1 & \beta_1^1 & \ldots & \beta_K^1 & \alpha_1^1\beta_1^1 & \ldots & \alpha_K^1
 &\beta_K^1
 \\
 \vdots & &&&\ldots&&&&\vdots\\
 \alpha_1^n & \ldots & \alpha_K^n & \beta_1^n & \ldots & \beta_K^n & \alpha_1^n\beta_1^n & \ldots & \alpha_K^n
 &\beta_K^n
 \\
 \end{pmatrix} \qquad \mbox{dim}(X) = n\times (2K + K^2)
 \]
 \[
  \mbox{cov}(\varepsilon_i,\varepsilon_j)=\rho(s_i-s_j)
\]

In order to estimate the regression and the correlation function
parameters we proceed by Quasi Generalized Least Squares: a
preliminary estimation of $\phi$ is given by Ordinary Least Squares,
$\phi^* = (X^tX)^{-1}X^tY$, the correlation function is estimated
from the residuals $\widehat{\varepsilon}=Y-X\phi^*$ and the final
estimate of $\phi$ is given by plugging the estimated correlation
matrix $\widehat{\Sigma}$ in the Generalized Least Squares
 formula $\widehat{\phi} =
(X^t\widehat{\Sigma}^{-1}X)^{-1}X^t\widehat{\Sigma}^{-1}Y$. If both
estimations of $\phi$ and $\Sigma$ are convergent and assuming
normal distribution of the residuals then \cite{Guyon}:

\[
  \sqrt{n}(\widehat{\phi}-\phi) \to \mathcal{N}(0,\lim_{n\to \infty}n(X^t\Sigma^{-1}X)^{-1})
\]
The estimation of $\Sigma$ is convergent under mild conditions
\cite{Cressie1} and the convergence of $\phi$ is assessed
for example when the functions are expanded on a splines basis
\cite{Cardot} or on a Karhunen expansion \cite{Muller}.

Significance of the predictors can be tested if the residuals are
assumed to be Gaussian, within the classical framework of linear
regression models.

Several parameters need to be set. The first choice is that of the
orthonormal basis. It can be  Fourier, splines, orthogonal
polynomials, wavelets. Then the order of truncation has to be
determined. The spatial correlation function of the residuals may be
of parametric form (exponential, Gaussian, spherical etc.). These
choices will be made by minimizing a cross validation criterion: a
sample with no missing data for all variables is determined, and for
each site of the sample a prediction of the diversity is computed
according to parameters estimated without the site in the sample.
The global criterion is the quadratic mean of the prediction error.

\section{Results}
\begin{figure}[H]
\includegraphics[width=9cm,height=16cm,angle=270]{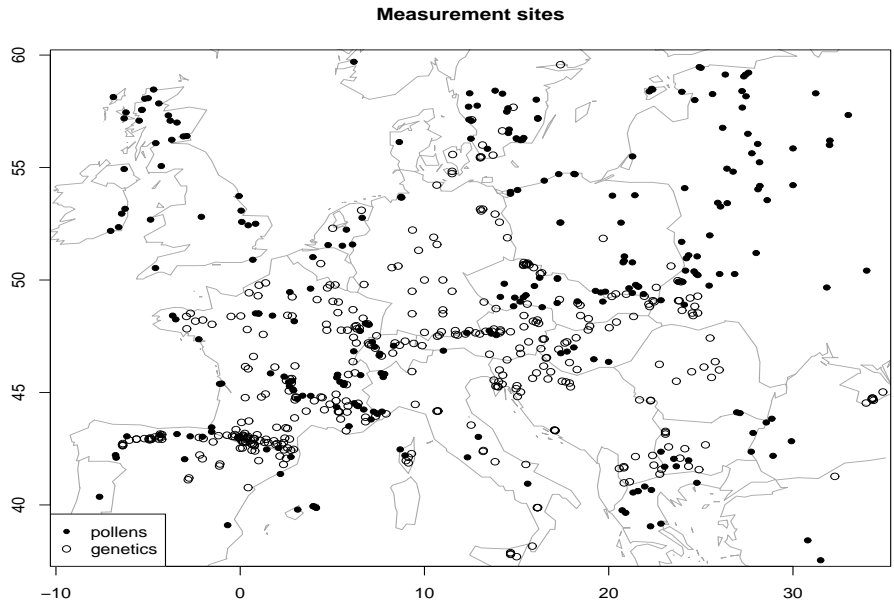}
\caption{\label{FigLoc}Locations of pollen (black dots) and genetic
(open circles) records. }
\end{figure}

Pollen was collected throughout Europe providing temporal estimation
of temperatures and precipitation. These estimations are not
regularly spaced, and have very different ranges from 1 Kyears to 15
Kyears. Beech genetic indices are recorded in forests and do not
coincide with the pollen locations. Figure \ref{FigLoc} shows the
locations of the two datasets.

Climate variables are continuous all over Europe but beech forests
have specific locations. In order to make our data to spatially
coincide, temperature and precipitation curves are firstly estimated
on a regular grid of time from 15 Kyears to present on sites where
are collected the genetic measures. 15 Kyears corresponds to  the
beginning of migration of plants  onto areas made free by the
retreating ice sheets.

\begin{figure}[H]
\includegraphics[width=9cm,height=16cm,angle=270]{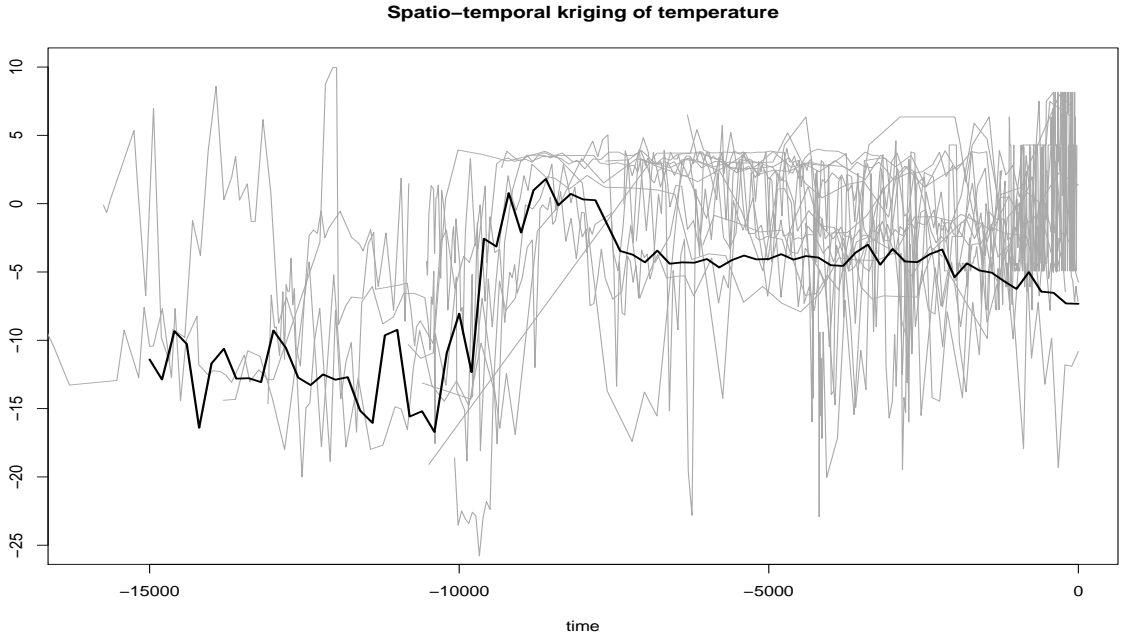}
\caption{\label{FigKrig} Resulting temperature curve (thick black
curve) from spatio-temporal kriging of 20 neighboring temperatures
curves from recorded pollen.}
\end{figure}

The interpolation is done by a spatio-temporal kriging assuming the
covariance function is exponential and separable. Figure
\ref{FigKrig} shows for a particular site the estimated temperature
curve together with some neighboring curves issued from collected
pollen.

\begin{figure}[H]
\includegraphics[width=9cm,height=16cm,angle=270]{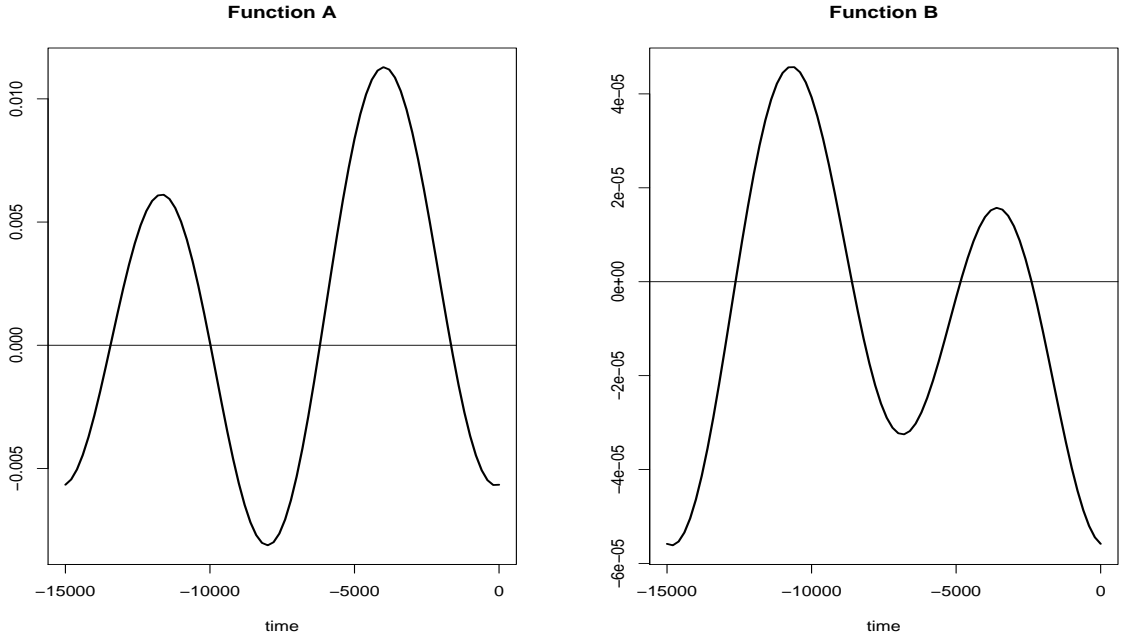}
\caption{\label{FigAB}Coefficient function $A$ of the temperature
and coefficient function $B$ of the precipitation}
\end{figure}

\begin{figure}[H]
\includegraphics[width=9cm,height=16cm,angle=270]{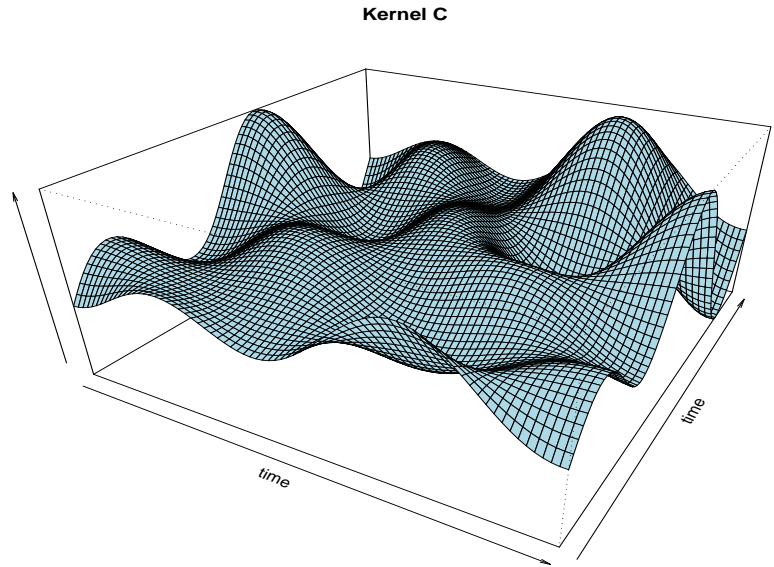}
\caption{\label{FigC} Kernel $C$ of the interaction
temperature-precipitation}
\end{figure}

We aim to predict genetic diversity with precipitation and
temperature curves. This corresponds to a functional regression
model with genetic diversity as dependent variable and temperature
and precipitation curves as predictor variable. The cross validation
criterion gives better results with an expansion of the predictor
variables on a Fourier basis of order 5. Figures \ref{FigAB} and
\ref{FigC} show the coefficient functions $A$, $B$, and kernel $C$.

The shape of the coefficient function $A$ shows that the term
$\langle A, \theta \rangle$ will be higher when the gap between
periods  before 7.5 Kyears and after 7.5 Kyears is higher
(temperatures before 7.5 Kyears are mostly negative), meanwhile the
shape of the coefficient function $B$ shows that the  term $\langle
B, \pi \rangle$ will be higher when the precipitation before 7.5
Kyears is higher (precipitation is positive). The  surface of kernel
$C$ is obviously not the product of two curves in the two
coordinates, showing an effect of interaction.

\begin{figure}[H]
\begin{center}
\includegraphics[width=6cm,height=10cm,angle=270]{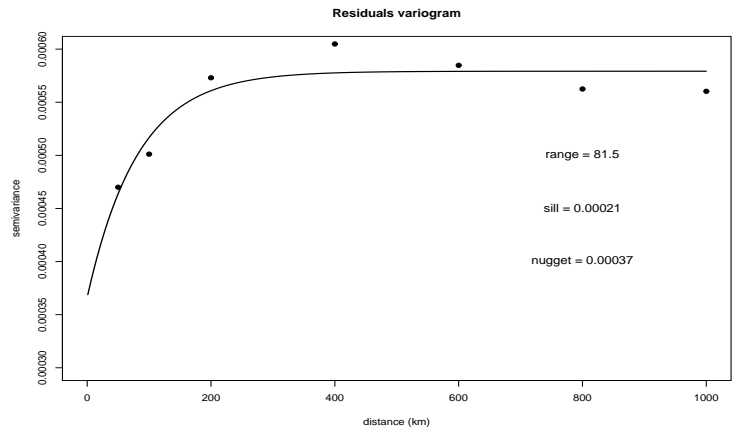}
\caption{\label{FigVario} Empirical and fitted variogram on the
residuals. }
\end{center}
\end{figure}

In Figure \ref{FigVario} the residual variogram  graph exhibits some
spatial dependence. An exponential variogram is fitted, and the
resulting covariance matrix is plugged into the GLS formula to
update the coefficients and test the effects of the temperature,
precipitation and interaction.

\begin{figure}[H]
\includegraphics[width=7cm,height=7cm,angle=270]{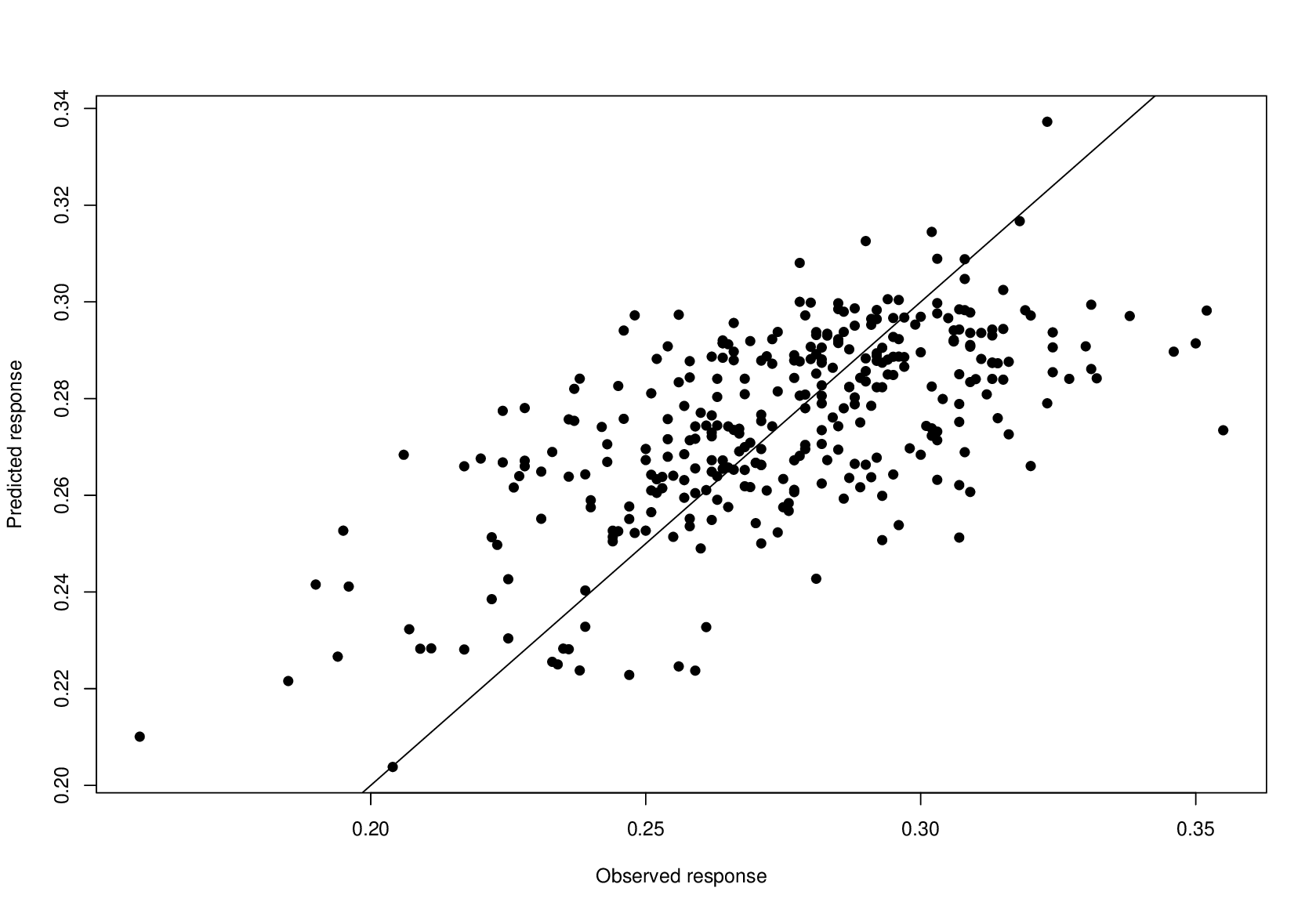}
\includegraphics[width=7cm,height=7cm,angle=270]{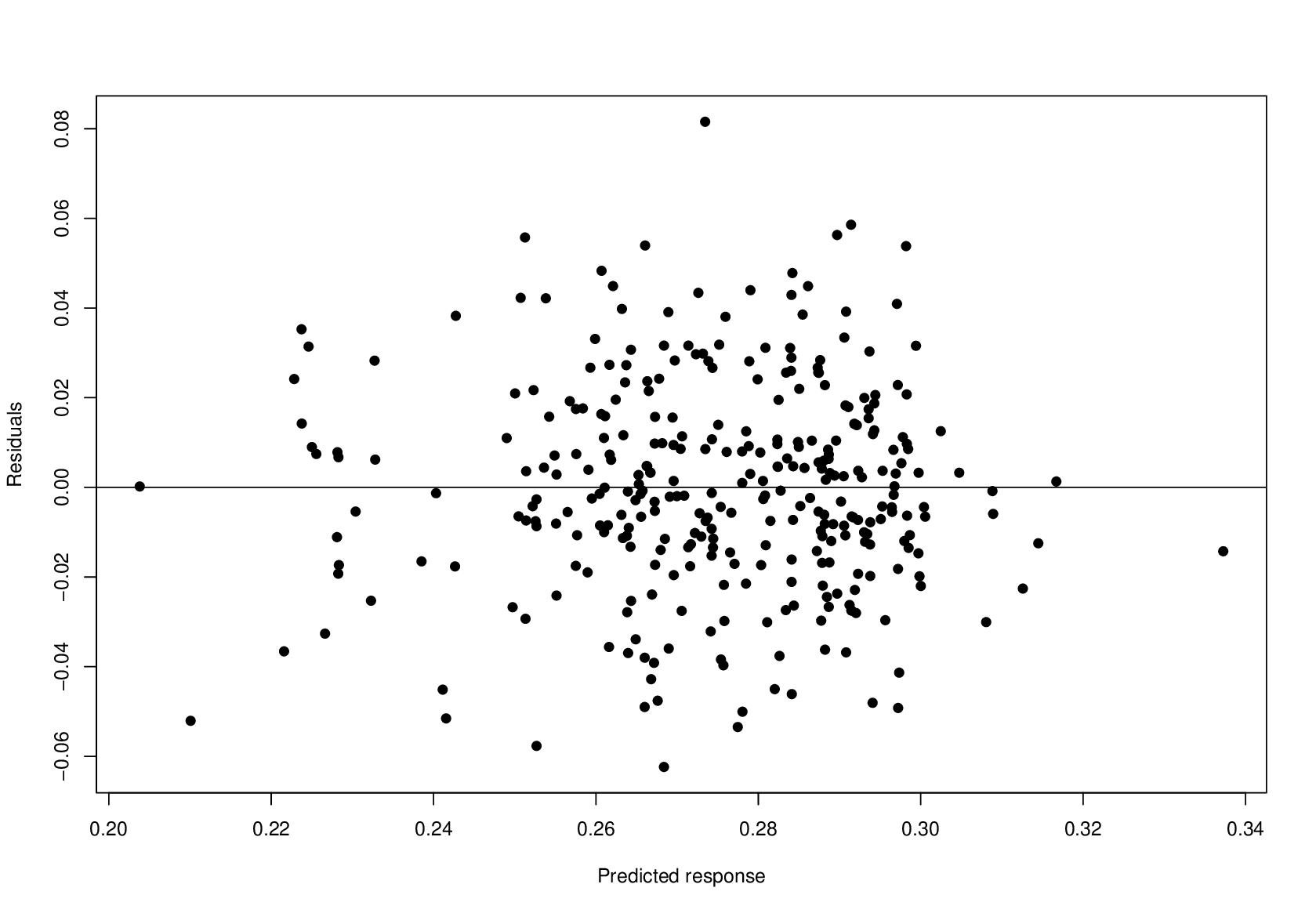}
\caption{\label{FigPredRes} Observed-Predicted response and
Predicted-Residuals graphs. }
\end{figure}

The graphs in Figure \ref{FigPredRes} show that the model explains a
part of the diversity variability. However it is far from explaining
 all the variability as the $R^2$ is equal to 0.31.

Table \ref{TabAnova} gives the analysis of variance of the four
nested models:
\begin{description}
\item[Model 1:] $\mathbb{E}(Y) = \mu+ \langle A ; \theta\rangle + \langle B ; \pi\rangle+ \langle C\theta ; \pi \rangle$
\item[Model 2:] $\mathbb{E}(Y) = \mu+ \langle A ; \theta\rangle + \langle B ; \pi\rangle$
\item[Model 3:] $\mathbb{E}(Y) = \mu+ \langle A ; \theta\rangle$
\item[Model 4:] $\mathbb{E}(Y) = \mu+ \langle B ; \pi\rangle$
\end{description}

\begin{table}[H]
\caption{Analysis of variance models of nested models\label{TabAnova} }
%\begin{tabular}{ccccccc}
%  &Res.Df  &   RSS  &Df &Sum of Sq  &      $F$  &  Pr($>F$)\\
%\hline
%Model 4  &  319 & 5911.8 &  5  &  9864.9 &127.2655 & $< $2.2e-16 ***\\
%Model 3  &  324 &15776.6 & 5 &  9864.9 &127.2655 &$< $2.2e-16 ***\\
%Model 2  &  319 & 5911.8 &25 &  1353.9 & 3.4934 &1.430e-07 ***\\
%Model 1   &  294 & 4557.8&&&&
%\end{tabular}
\end{table}

The $p$-values (2.2e-16) of the tests H$_0$: model 3 (model 4) against H$_1$:
model 2 and (1.430e-07) of the test H$_0$: model 2 against H$_1$:
model 1 show that the interaction and the two variables have a
strong effect.

\begin{figure}[H]
\includegraphics[width=9cm,height=16cm,angle=270]{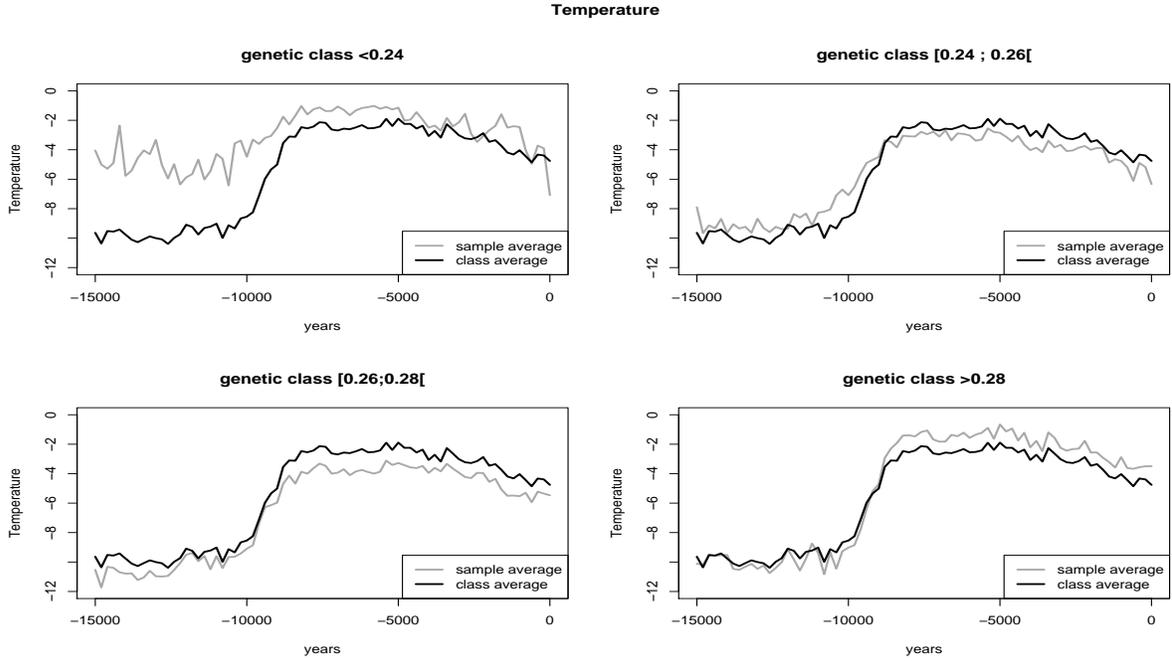}
\caption{\label{FigTemp}Pattern of temperature curves according to
the range of the predicted response. }
\end{figure}

\begin{figure}[H]
\includegraphics[width=9cm,height=16cm,angle=270]{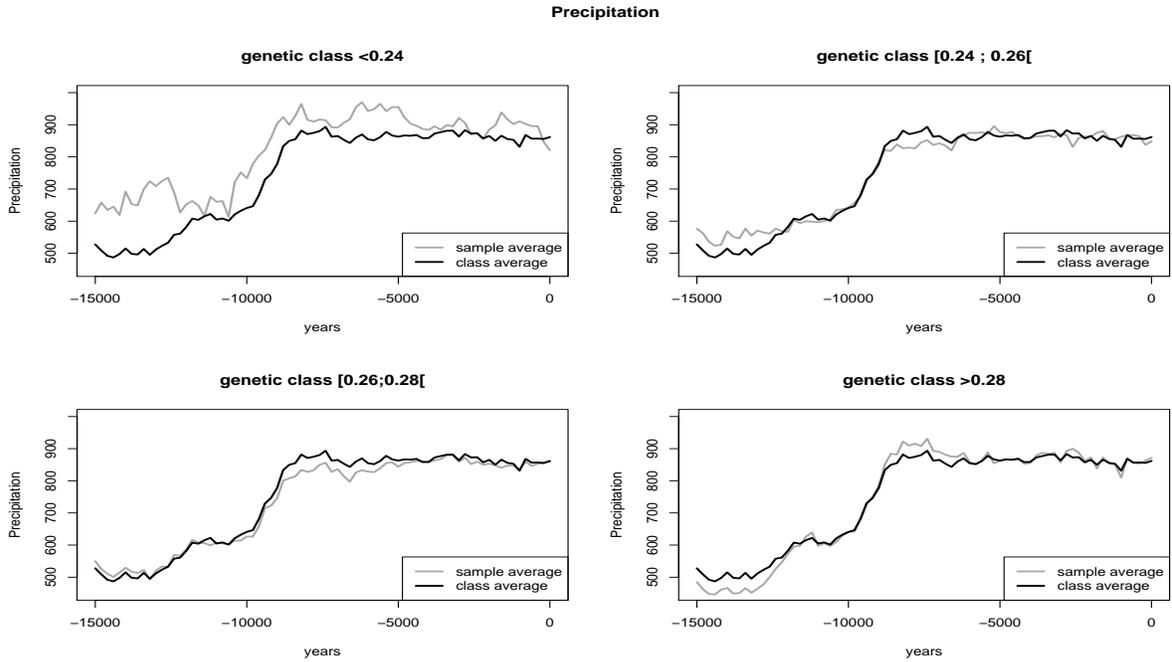}
\caption{\label{FigPrec} Pattern of precipitation curves according
to the range of the predicted response. }
\end{figure}

To give a better understanding of the regression model we divide the
predicted response range into 4 classes: less than 0.24,
]0.24;0.26], ]0.26;0.28] and greater than 0.28. Figures
\ref{FigTemp} and \ref{FigPrec} show the shapes of temperature and
precipitation curves for each class. When low ($<$ 0.24) diversity
is predicted, temperature curves are globally higher than the
averaged temperature curves on all the sample. As the predicted
diversity becomes higher the gap between the two periods before 7.5
Kyears and after 7.5 Kyears gets more pronounced. This effect is
less evident for precipitation, low diversity is predicted when the
precipitation is higher on the first period than the averaged
precipitation curves on all the sample. When the predicted diversity
is higher than 0.24 there seems to be  no effect of precipitation on
its the level.

When the change of climate during the Holocene (12 Kyears to
present) is significant the diversity is higher. This mostly
concerns northern and western Europe. This is coherent with previous
studies \cite{cheddadi}. After 12 Kyears and throughout the Holocene
the climate was no longer uniform all over Europe. The largest
mismatch between NW and SE Europe occurred around 9 Kyears and 5
Kyears. By 5 Kyears, all deciduous tree taxa (such as beech) were
outside their glacial refugia.

\section{Conclusion}
The classical linear functional model has been extended in a
straightforward manner to the case of two functional predictors with
an interaction term, and with spatially correlated residuals. Such a
model applied to complex paleoecological and biodiversity data
emphasizes an interesting relationship between climate change and
genetic diversity: diversity is higher when the change in climate
(mostly temperature) during the Holocene (12 Kyears to present) was
sizeable and lower when temperature and precipitation are both
globally higher over the whole period. This model may be improved in
several ways. The spatial effect may be handled in other ways, by
 means of a mixed structure or with other kinds of
correlation matrix structure. In this first attempt we have
neglected the random structure and the correlation of the
predictors. Taking into account these two characteristics should
give a better way to understand the real effect of climate on
biodiversity.

\clearpage

\bibliographystyle{plain}

\end{document}